\documentclass[a4paper,11pt]{article}

\pdfoutput=1 
\usepackage{jcappub}

\usepackage{graphicx}
\usepackage{longtable}
\usepackage{float}
\usepackage{dcolumn}
\usepackage{bm}
\usepackage{appendix}
\usepackage{multirow}
\usepackage{color}
\usepackage[utf8]{inputenc}
\usepackage{footmisc}
\usepackage{txfonts}
\usepackage{xcolor}
\usepackage{graphicx}
\usepackage{bm}
\usepackage{ulem}
\usepackage{multirow, array, float, tabularx}
\usepackage{amsmath}
\usepackage[colorlinks=true,linkcolor=blue,citecolor=blue]{hyperref}

\newcommand{\be}{\begin{equation}}
\newcommand{\ee}{\end{equation}}
\newcommand{\bea}{\begin{eqnarray}}
\newcommand{\eea}{\end{eqnarray}}
\newcommand{\bes}{\begin{subequations}}
\newcommand{\ees}{\end{subequations}}

\newcommand{\bc}{\begin{center}}
\newcommand{\ec}{\end{center}}

\usepackage{wasysym}
 
\begin{document}
 
\title{Discrepancy between cosmological and electroweak observables in Higgs Inflation}

\author[a]{Jamerson G. Rodrigues}\emailAdd{jamersoncg@gmail.com}

\author[b,c,d]{Micol Benetti}\emailAdd{micol.benetti@unina.it}

\author[e]{Jailson S. Alcaniz}\emailAdd{alcaniz@on.br}

\affiliation[a]{Departamento de F\'{\i}sica, Universidade Federal do Rio Grande do Norte, 59078-970, Natal, RN, Brasil}

  \affiliation[b]{Dipartimento di Fisica "E. Pancini", Universit\`{a} di Napoli  ``Federico II'', Complesso Universitario di Monte Sant'Angelo,
  Edificio G, Via Cinthia, I-80126, Napoli, Italy}

  \affiliation[c]{Istituto Nazionale di Fisica Nucleare (INFN) Sezione
  di Napoli, Complesso Universitario di Monte Sant'Angelo, Edificio G,
  Via Cinthia, I-80126, Napoli, Italy}
  
  \affiliation[d]{Scuola Superiore Meridionale, Universit\`{a} di Napoli ``Federico II'', Largo San Marcellino 10, 80138 Napoli, Italy}
  
\affiliation[e]{Departamento de Astronomia, Observat\'orio Nacional, 20921-400 Rio de Janeiro, RJ, Brasil}

  \abstract{In this work, we revisit the non-minimally coupled Higgs Inflation scenario and investigate its observational viability in light of the current Cosmic Microwave Background, Baryon Acoustic Oscillation and type Ia Supernovae data. We explore the effects of the Coleman-Weinberg approximation to the Higgs potential in the primordial universe, connecting the predictions for the Lagrangian parameters at inflationary scales to the electroweak observables through Renormalization Group methods at two-loop order. As the main result, we find that observations on the electroweak scale are in disagreement with the constraints obtained from the cosmological data sets used in the analysis. Specifically, an $\approx 8\sigma$-discrepancy between the inflationary parameters and the electroweak value of the top quark mass is found, which suggests that a significant deviation from the scenario analysed is required by the cosmological data.}

\maketitle

\section{Introduction}
\label{Introduction}

The inflationary paradigm is now consolidated as the standard  description of the primordial universe~\cite{Guth:1980zm,Linde:1981mu,Albrecht:1982wi,Mukhanov:1981xt,Guth:1982ec,Starobinsky:1982ee,Hawking:1982cz}. However, in spite of its observational successes, the very nature of the inflaton field remains unknown, and a large number of  inflationary models has been proposed since the conception of the theory \cite{Martin:2013tda}. In particular, the model proposed by Bezrukov and Shaposhnikov \cite{Bezrukov:2007ep} offers one of the most promising alternatives, given its simplicity and ability to describe the current data \cite{Aghanim:2018eyx}. In such a model inflation is driven by the only scalar field observed so far, the Higgs boson \cite{Aad:2012tfa,Chatrchyan:2012xdj}, and is known as Higgs Inflation.

In the well--known slow-roll approximation, in order to induce cosmic acceleration, the dynamical equations for the inflaton field  enables a slowly varying solution, mimetizing a cosmological constant. In the Higgs Inflation scenario,  the inﬂaton potential can exhibit a plateau at high energies which is achieved through the introduction of a non-minimal coupling between the Higgs field and the Ricci scalar. Although an appealing approach, the introduction of a non-minimal interaction between the inflaton and gravity may potentially compromise the quantum coherency of the theory. In particular, the authors in Refs.~\cite{Burgess:2009ea,Barbon:2009ya,Burgess:2010zq,Hertzberg:2010dc} discuss the lost of quantum unitarity at the scale of energy $\Lambda = M_P/\xi$, which is far bellow the inflationary regime $h>M_P/\sqrt{\xi}$. We refer the reader to \cite{Lerner:2009na,Bezrukov:2010jz,Barvinsky:2009ii,Bezrukov:2013fka,Rubio:2018ogq} for a different interpretation. 

In addition, the observational feasibility of the model rely on the stability of the Standard Model of Fundamental Particles (SM) up to inflationary energy scales. However, following the Renormalization Group Equations (RGE) of the SM couplings, one obtains that the Higgs quartic coupling evolve to small negative values \cite{Espinosa:2007qp,Bezrukov:2012sa,Degrassi:2012ry,Buttazzo:2013uya,Bednyakov:2015sca}, which amounts to saying that the SM develops a unstable scalar potential for energy scales larger than a critical value around $10^{10}-10^{11}$ GeV.  Naturally, such behaviour compromises  the observational feasibility of the Higgs Inflation. Nevertheless, it is particularly interesting that the SM phase diagram sits right at the edge of stability \cite{Froggatt:1995rt,Shaposhnikov:2009pv,Bezrukov:2012sa,Degrassi:2012ry,EliasMiro:2012ay,Buttazzo:2013uya,Masina:2012tz,Salvio:2016mvj,Domenech:2020yjf}. In this sense, the magnitude of the Higgs quartic coupling may be an important hint for new physics. 


In this work we revisit the Higgs inflationary model in its minimal content. In particular, we perform a detailed investigation of the observational viability of the Coleman-Weinberg approximation to the one-loop effective Higgs potential in light of the most recent Cosmic Microwave Background (CMB), Baryon Acoustic Oscillation (BAO) and Supernova data \cite{Aghanim:2019ame,Ade:2018gkx,Beutler:2011hx,Alam:2016hwk,Scolnic:2017caz}. Applying a Monte Carlo Markov Chains (MCMC) parameter estimation, we impose constraints on the radiative corrections to the Higgs quartic coupling at inflationary energy scale. Moreover, we also solve the two-loop Renormalization Group Equations (RGE) for the SM parameters to obtain the corresponding constraints at the electroweak scale and the upper limit to the top quark pole mass, in order to examine if current cosmological data are compatible with the electroweak phenomenology  and the inflationary dynamics of the Higgs field. Our analysis shows a large discrepancy between the value of the top quark pole mass required by cosmological observations and the one measured by electroweak experiments, $M_t = 172.76 \pm 0.30$ GeV \cite{Zyla:2020zbs}

This paper is organized as follows: in Sec. \ref{sec: Potential} we present the effective potential employed in our analyses. In Sec. \ref{sec:Higgs}, we introduce the Higgs Inflation scenario whereas in Sec. \ref{sec:Slow-Roll} we perform the slow-roll analysis for the model. We discuss the Renormalization Group approach assumed in our study in Sec. \ref{sec:RGE}. In Sec. \ref{sec:Method&Results}, we present the method of analysis to test the theory with cosmological data and discuss our analysis results.  The main conclusions of this work are presented in Sec. \ref{sec:Conclusions}.

\section{Effective Potential}
\label{sec: Potential}

Before considering the Higgs Inflation mechanism, let's discuss the effective potential of the Standard Model minimally coupled to gravity. 
For the Higgs field, the RGE-improved effective potential can be written in the approximated form
\begin{equation}
    V_{eff}(h) = \lambda_{eff}(\mu) \left( \lvert \mathcal{H}\rvert^2 - \frac{v^2}{2} \right)^2 \approx \frac{\lambda_{eff}(\mu)}{4}h^4,
\end{equation}
where $\mathcal{H} = \left(0, (h+v)/\sqrt{2}\right)^T$ is the Higgs doublet, $v$ is the standard model vacuum expectation value and $\mu$ is the renormalization scale \cite{Casas:1994qy}. The effective coupling $\lambda_{eff}(\mu)$ encodes the contributions from the relevant running couplings. For a perturbative theory, it can be expanded as a tree level component plus the loop contributions,
\begin{equation}
\lambda_{eff}(\mu) = e^{4\Gamma(\mu)}\left[\lambda(\mu) + \lambda^{(1)}(\mu) + \lambda^{(2)}(\mu) + \ldots \right], \label{Exper}
\end{equation}
with the ellipsis representing high order terms, whereas the factor $\Gamma(\mu)$ takes into account the field strength renormalization,
\begin{equation}
    \Gamma(\mu) \equiv \int_{\mu_0}^{\mu_1}\gamma(\mu)d\ln{\mu},\label{eq:FReescaling}
\end{equation}
and $\gamma(\mu)$ is the Higgs anomalous coupling.

The two-loop effective potential for the SM was first derived by Ford et al. in \cite{Ford:1992pn}. Latter, the authors of \cite{Degrassi:2012ry} introduced a compact form for the effective potential in the limit of $\lambda \rightarrow 0$, subsequently enhancing the accuracy of the analysis in \cite{Buttazzo:2013uya}. In particular, the one-loop contributions to the effective coupling recover the structure derived by Coleman and Weinberg \cite{Coleman:1973jx},
\begin{equation}
    \lambda^{(1)}(\mu) = \frac{1}{(4\pi)^2}\sum_p N_p \kappa^2_p(\mu)\left( \ln{\frac{\kappa_p(\mu)e^{2\Gamma(\mu)}h^2}{\mu^2}}-C_p\right),\label{Lamb1}
\end{equation}
where $p$ runs for all the species of particles contributing to the loop diagrams, $N_p$ accounts for the degrees of freedom, $\kappa$ rises from the field-dependent mass squared and $C_p$ is a renormalization scheme dependent constant. The predominant contributions for the summation in \eqref{Lamb1} are the ones coming from the top quark $t$, the weak gauge bosons $W$ and $Z$, the Higgs $h$ and Goldstone $\chi$ bosons loops. Table \ref{coeff} summarizes the values for these coefficients for the $\overline{\text{MS}}$ renormalization scheme and the Landau gauge. 

\begin{table}[h!]
\vskip .5cm 
\centering

\begin{tabular}{  r | c  c  c   c  c      }
 $p$ & $t$ & $W$ & $Z$ & $\phi$  & $\chi$  \\
\hline
$N_p$         & $-12$ & $6$ & $3$  & $1$ & $3$   \\
$C_p$         &  $3/2$ & $5/6$  & $5/6$ & $3/2$ & $3/2$  \\
$\kappa_p$ &  $y^2_t/2$ & $g^2/4$ & $(g^2+g'^2)/4$ & $3\lambda$ & $\lambda$
\end{tabular}
\caption{Coefficients for the effective quartic coupling in eq. (\ref{Lamb1}). }
\label{coeff}
\vskip .5cm 
\end{table}

Despite the increasing accuracy obtained in the extrapolation of the Standard Model properties \cite{Mihaila:2012fm,Mihaila:2012pz,Chetyrkin:2012rz,Zoller:2015tha,Bednyakov:2015ooa}, the high-energy behaviour of the theory seems to be fairly well approximated by the leading-order terms in expression \eqref{Exper}. As discussed in \cite{Degrassi:2012ry,Buttazzo:2013uya,Holthausen:2011aa,EliasMiro:2011aa}, the Higgs quartic coupling $\lambda$ and its $\beta$-function $\beta_\lambda$ run to small values, reaching a minimum somewhere bellow the Planck scale. This suggests that $\lambda$, as well as its $\beta$-function, holds only a weak dependence with the renormalization scale $\mu$. One might assume $\beta_\lambda$ to be constant, yielding the familiar Coleman-Weinberg form for the effective Higgs potential \cite{Coleman:1973jx,Sher:1988mj,Okada:2010jf},
\begin{equation}
    V_{eff}(h) \approx \left(\frac{\lambda(M)}{4} + \frac{\beta_\lambda (M)}{4}\ln{\frac{h}{M}}\right)h^4, \label{veff}
\end{equation}
where $M$ is some high-energy scale. For simplicity, the constant terms in \eqref{Lamb1} are redefined into an unobservable phase shift of the Higgs field $h$ and the anomalous coupling is assumed to vanish, $\gamma \sim 0$.

The potential energy in \eqref{veff} is appropriate to describe the standard model for field displacements close to $M$, otherwise spoiling the perturbative aspect of Higgs-Higgs scattering processes. In particular, the problem with large-logarithms is avoided as long as $\beta_\lambda (M)/4\ln{(h/M)}\ll 1$.

\section{Higgs Inflation}
\label{sec:Higgs}

Although the application of modified gravity theories in order to achieve a slow-roll phase in the early universe is not particularly recent~\cite{Starobinsky:1980te,Salopek:1988qh}, the viability of a Higgs driven inflation in such scenarios has been widely debated in the last decades \cite{Bezrukov:2007ep,Barvinsky:2008ia,GarciaBellido:2008ab,Bezrukov:2014bra,Lee:2018esk}. Essentially, the model proposed by Bezrukov and Shaposhnikov \cite{Bezrukov:2007ep} expands the canonical Einstein-Hilbert sector with a non-minimal coupling between the Higgs field and the Ricci scalar. Such configuration produces a inflationary plateau at large field regime, driving the model predictions to the sweet-spot of the CMB observations \cite{Aghanim:2018eyx}.

In the Higgs Inflation, the expanded gravity sector is defined in the Jordan frame, where the inflationary Lagrangian assumes the form,
\begin{equation}
    {\cal L} = \frac{1}{2} (\partial_\mu h)^{\dagger}(\partial^\mu h)-\frac{M_P^2R}{2}-\frac{1}{2}\xi {h}^2 R -V_J(h),
 \label{Ljordan}
\end{equation}
where $M_P=2.435\times 10^{18}$ GeV is the reduced Planck mass. Here, the non-minimal interaction is parameterized by the dimensionless coupling $\xi$.

In practical sense, it is useful to recover the canonical Einstein-Hilbert gravity in order to compute the inflationary parameters. To this purpose, one can perform a set of conformal transformations in the metric 
\cite{Birrell:1982ix,Accioly:1993kc,Faraoni:1998qx}:
\be
\tilde g_{\mu \nu}=\Omega^2g_{\mu \nu}\,\,\,\,\,\,\mbox{where} \,\,\,\,\,  \Omega^2= 1+\frac{\xi {h}^2}{M^2_P},
\label{conformaltransf}
\ee
which makes the kinetic energy of the inflaton field non-canonical. The process is finished by the field redefinition
\be
\chi^\prime \equiv \frac{d\chi}{dh}=\sqrt{\frac{\Omega^2 +6\xi^2 {h}^2/M_P^2}{\Omega^4}}.\label{InflatonRed}
\ee
Finally, the Lagrangian with minimal gravity sector and canonical kinetic term is defined in the Einstein frame,
\be
 {\cal L} = -\frac{M^2_{P} \tilde R}{2}+\frac{1}{2} (\partial_\mu \chi)^{\dagger}(\partial^\mu \chi)-V(\chi)\,,
 \label{LEinstein}
\ee
where $V(\chi)=\frac{1}{\Omega^4}V_J(h\left[\chi\right])$.


The scalar potential in \eqref{LEinstein} is the one employed to the inflationary parameter estimation. In the original proposal \cite{Bezrukov:2007ep}, the authors performed the analysis of the model at tree level, obtaining for the spectral index and the tensor-to-scalar ratio the values $n_S\simeq 0.97$ and $r \simeq 0.0033$, thus in excellent agreement with current Planck data \cite{Aghanim:2018eyx}. Also, the measured value of the amplitude of scalar perturbation, $A_S \simeq 2.1 \times 10^{-9}$, impose a strong constraint to the non-minimal coupling $\xi \sim 10^4$.

Latter, the quantum corrections to the inflaton potential were probed to play a major role at inflationary dynamics \cite{Barvinsky:2008ia,Barvinsky:1998rn,Bezrukov:2008ej,DeSimone:2008ei,Barvinsky:2009fy}. There are, however, several dubieties in what concerns quantum effects in the model. The unnaturally large value of $\xi$ supposedly give rise to a new scale, $\Lambda =M_P/\xi$, associated to the loss of quantum unitarity of the system \cite{Burgess:2009ea,Barbon:2009ya,Burgess:2010zq,Hertzberg:2010dc} (see \cite{Lerner:2009na,Bezrukov:2013fka,Rubio:2018ogq} for a different point of view). In addition, there is an inherent ambiguity in the definition of effective potential of the inflaton. To compute the radiative corrections to such scenario one has to choose a frame of reference, leading to two nonequivalents results, namely prescription I \cite{Bezrukov:2007ep,Bezrukov:2008ej} (Einstein frame) and prescription II schemes \cite{Barvinsky:2008ia} (Jordan frame). The former presents an attractive alternative to investigate the high-energy behavior of the theory, where the effects of the non-minimal coupling are strong. The latter seems to be more appropriate to study the connection between the inflationary predictions and the electroweak scale observables, since the physical distances are measured in the Jordan frame \cite{Bezrukov:2008ej}. Without further knowledge about the ultraviolet completion of the model it is not clear whether frame is appropriate to compute the quantum effects.


At one-loop level, however, the differences in the two prescription schemes resume to a spurious relation between the renormalization scale and the gauge bosons background masses, $\mu(h)$ \cite{Masina:2018ejw,Allison:2013uaa}. In view of this, we opt to employ in the following sections the prescription II scheme, leaving the analysis of the prescription I scheme for a forthcoming communication.

\section{Slow-Roll Analysis}
\label{sec:Slow-Roll}
Before considering the Bayesian analysis of the Higgs Inflation model, it is useful to revisit the slow-roll predictions of the inflationary parameters. To this end, we compute the one-loop effective potential in light of the prescription II procedure. 

Since the radiative corrections to the tree level Lagrangian are computed in the Jordan frame, the resulting effective potential reduces to the familiar Coleman-Weinberg form presented in eq. \eqref{veff} \cite{Barvinsky:2008ia}. After the set of conformal transformations \eqref{conformaltransf} and the field redefinition \eqref{InflatonRed}, we have the Einstein frame description of the theory,
\bea
V(\chi) = && \frac{\lambda M^4_P}{4\xi^2}\left(1-\exp{\left(-\frac{2}{M_P}\sqrt{\frac{1}{6}}\chi \right)} \right)^2 \nonumber \\
&& \times \left( 1+a^\prime \ln{\left(\sqrt{\frac{1}{\xi}\exp{\left(\frac{2}{M_P}\sqrt{\frac{1}{6 }}\chi \right)}-\frac{1}{\xi}} \right)} \right). \label{eq:InfPot}
\eea
The large field regime is assumed, $\chi \gg \sqrt{6}M_P$, in order to obtain $h(\chi)$. Note that the deviation from the tree level potential is quantified by the parameter $a^\prime \equiv \beta_\lambda/\lambda$. All the couplings are computed at renormalization scale $M=M_P$.

Once with the effective scalar potential of the model, the analysis of its high-energy behaviour is straightforward. In particular, the study of the inflationary dynamics follows from the slow-roll parameters,
\begin{eqnarray}
 &\epsilon & = \frac{M^2_{P}}{2}\left(\frac{ V^{\prime}}{ V}\right)^2, \quad \quad
 \eta  = M^2_{P}\left( \frac{V^{\prime \prime}}{V } \right),
 \label{slowparameters}
\end{eqnarray}
where $^\prime$ indicates derivative with respect to the canonical field, $\chi$. The beginning of inflation takes place in the slow-roll regime $\epsilon,\eta \ll 1$ and continues until $\epsilon,\eta \simeq 1$. One can compute the predictions for the spectral index and the tensor-to-scalar ratio through,
\be
 n_S=1-6\epsilon+2\eta, \quad \quad \quad r=16\epsilon.
 \label{PParameters}
\ee

Another important observable quantity is the amplitude of scalar perturbations, which is related with the Primordial Power Spectrum ($P_R$) of curvature perturbations produced during inflation:
\begin{equation}
A_s =\left. P_R \right|_{k=k_{\ast}} =\left.\frac{V}{24M^4_P\pi^2\epsilon}\right|_{\chi=\chi_{\ast}}.
\label{eq:PR}
\end{equation}

All the aforementioned parameters are computed for the field strength $\chi_{\ast}$, associated to the energy density at which a chosen pivot scale, $k_\ast$, crossed the Hubble horizon. Although $\chi_{\ast}$ cannot be directly measured, it can be related to the amount of expansion the universe experienced, from the horizon crossing moment up to the end of inflation. Such quantity is defined by the number of e-folds,
\begin{equation}
    N_{\ast} = -\frac{1}{M^2_P}\int_{\chi_{\ast}}^{\chi_e}\frac{V}{V^\prime}d\chi. \label{efolds}
\end{equation}

Similarly to $r$, $n_S$ and $A_S$, the number of e-folds is not a free parameter, but an observable quantity associated to the evolution of cosmological scales \cite{Liddle:2003as}. Its main source of uncertainty lies from the lack of information about the reheating process, yielding to this observable a moderate model dependence. For the class of inflationary models with large non-minimal coupling, such as Higgs Inflation, the number of e-folds can be evaluated between $50-60$ for all relevant scales \cite{GarciaBellido:2008ab,Bezrukov:2008ut,Repond:2016sol,Almeida:2018oid,Gong:2015qha,Ballesteros:2016xej}.

With the value of the number of e-folds, one can use the expression \eqref{efolds} to obtain $\chi_{\ast}$, and then compute the model predictions for the inflationary parameters. 
Such a procedure is not possible analytically, unless under a set of reasonable assumptions \cite{Barvinsky:2008ia,Barvinsky:2009ii}. Here, we opt to employ numerical methods to solve the observable in terms of the free parameters $a^\prime$ and $\xi$. In particular, we set the number of e-folds to $N_{\ast} = 55$, in order to compare the model predictions with the Planck data.

\begin{figure}[!h]
\centering
\includegraphics[scale=0.35]{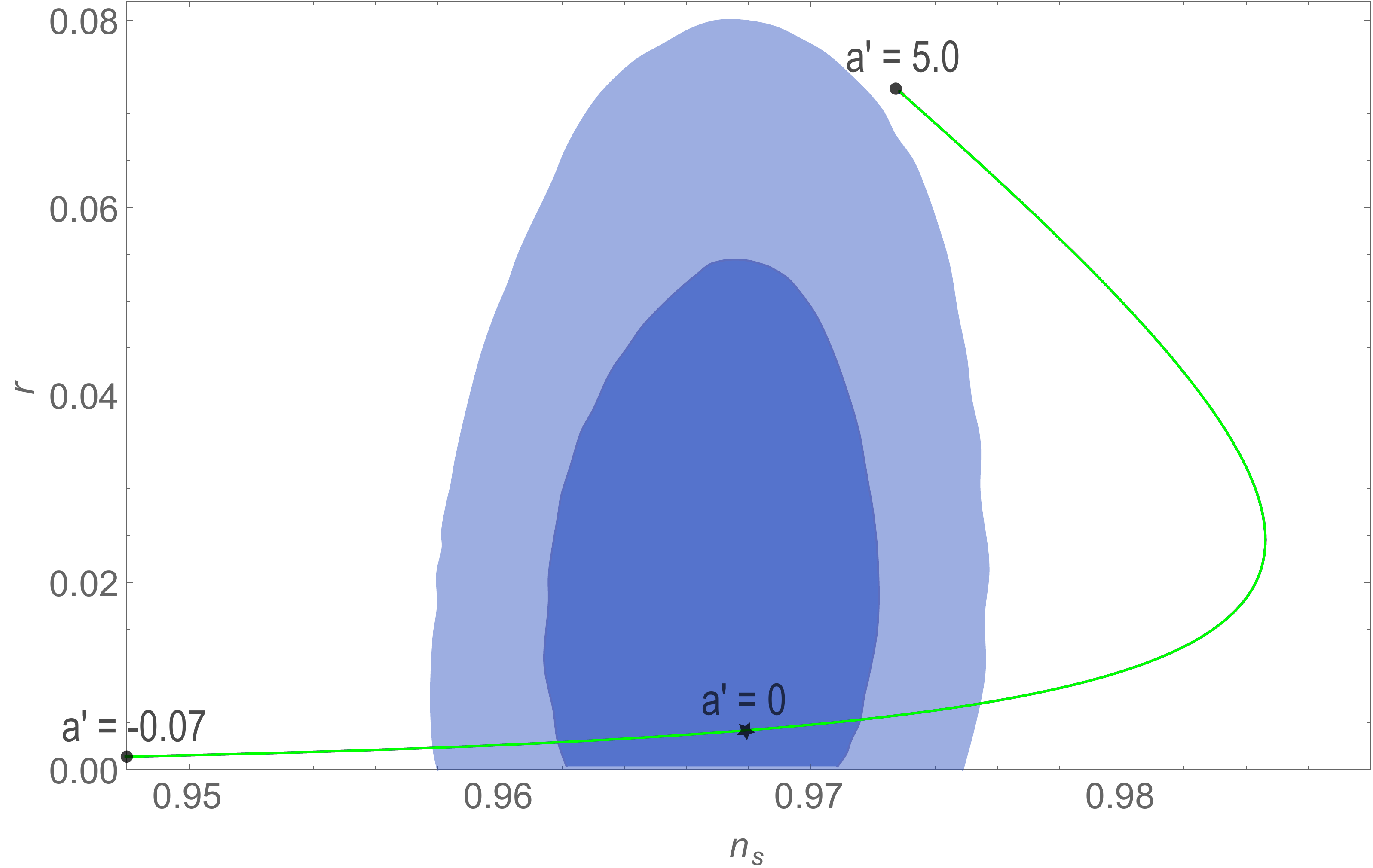}
\caption{$n_S$ vs $r$ for $\xi = 100,\,\, 1000 \text{ and } 5000$. From the inferior left extreme of the curve to its superior right, the parameter $a^\prime$ assumes the values $-0.07 \leq a^\prime \leq 5$. The blue areas show the favoured regions by Planck2018, with $68\%$ and $95\%$ confidence level (Planck $TT,TE,EE+lowE+lensing+BK14+BAO$ data set) \cite{Aghanim:2018eyx}.
}
\label{Fig:ns_r}
\end{figure}

Fig. \ref{Fig:ns_r} shows the predictions of the model in the $n_S \times r$ plane, for $\xi = 100, 1000, 5000$ and $-0.07 \leq a^\prime \leq 5$. The first aspect one shall note in the figure is the superposition of the curves for the different values of $\xi$, indicating that the inflationary observable is not sensitive to this parameter in the strong coupling regime. Such behaviour was pointed out in \cite{Rodrigues:2020dod}. On the other hand, $n_S$ and $r$ seems to be highly susceptible to variations in the radiative corrections. In particular, the predictions of the model converge to the Planck favoured region ($95$ C.L.) for $-0.02 \leq a^\prime \leq 0.28$, with the null correction case aligned with the $68\%$ C.L. Planck result ($n_S \simeq 0.965$ and $r \simeq 0.0035$ for $a^\prime = 0$) \cite{Aghanim:2018eyx}.

\begin{figure}[!h]
\centering
\includegraphics[scale=0.23]{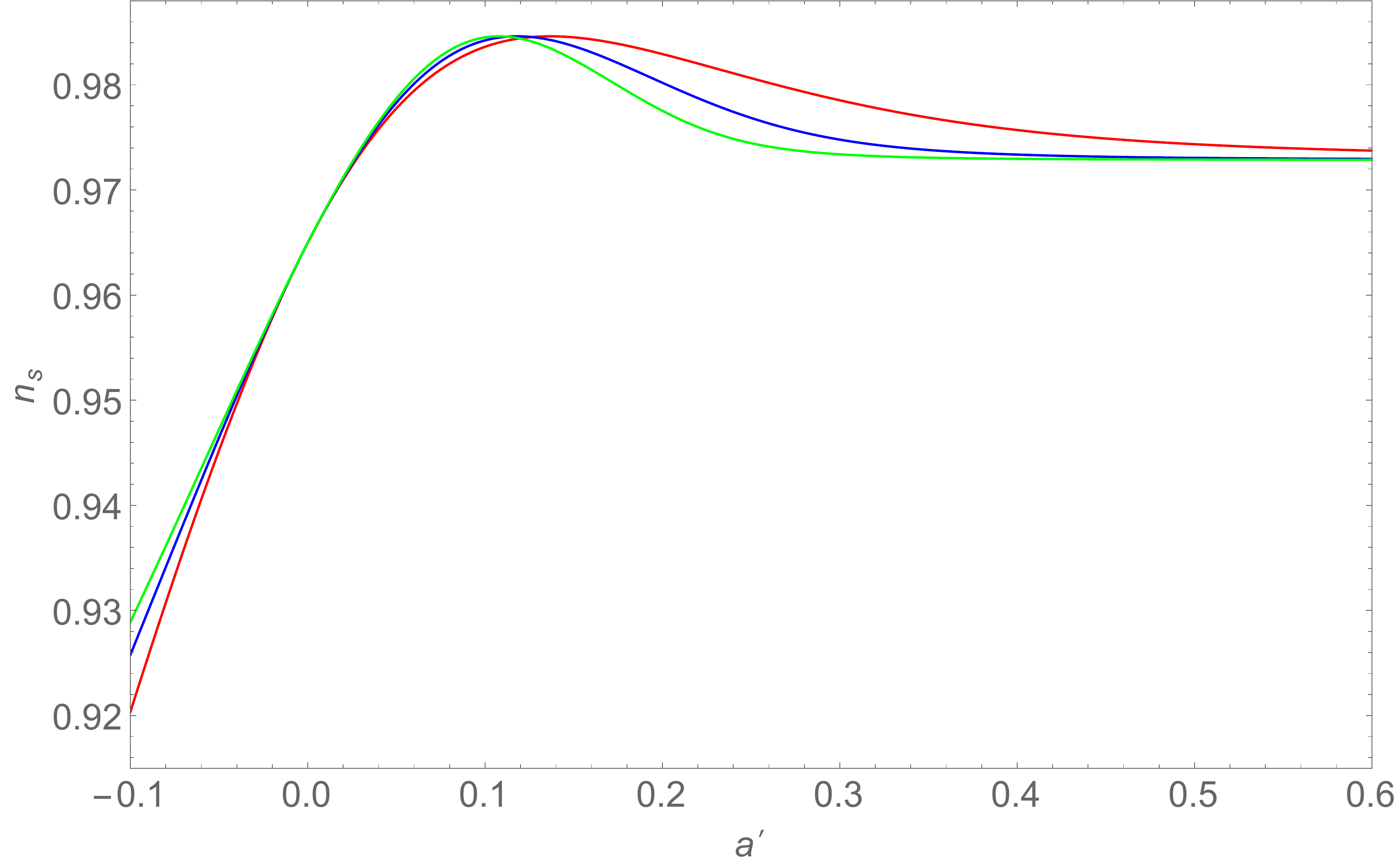}
\includegraphics[scale=0.23]{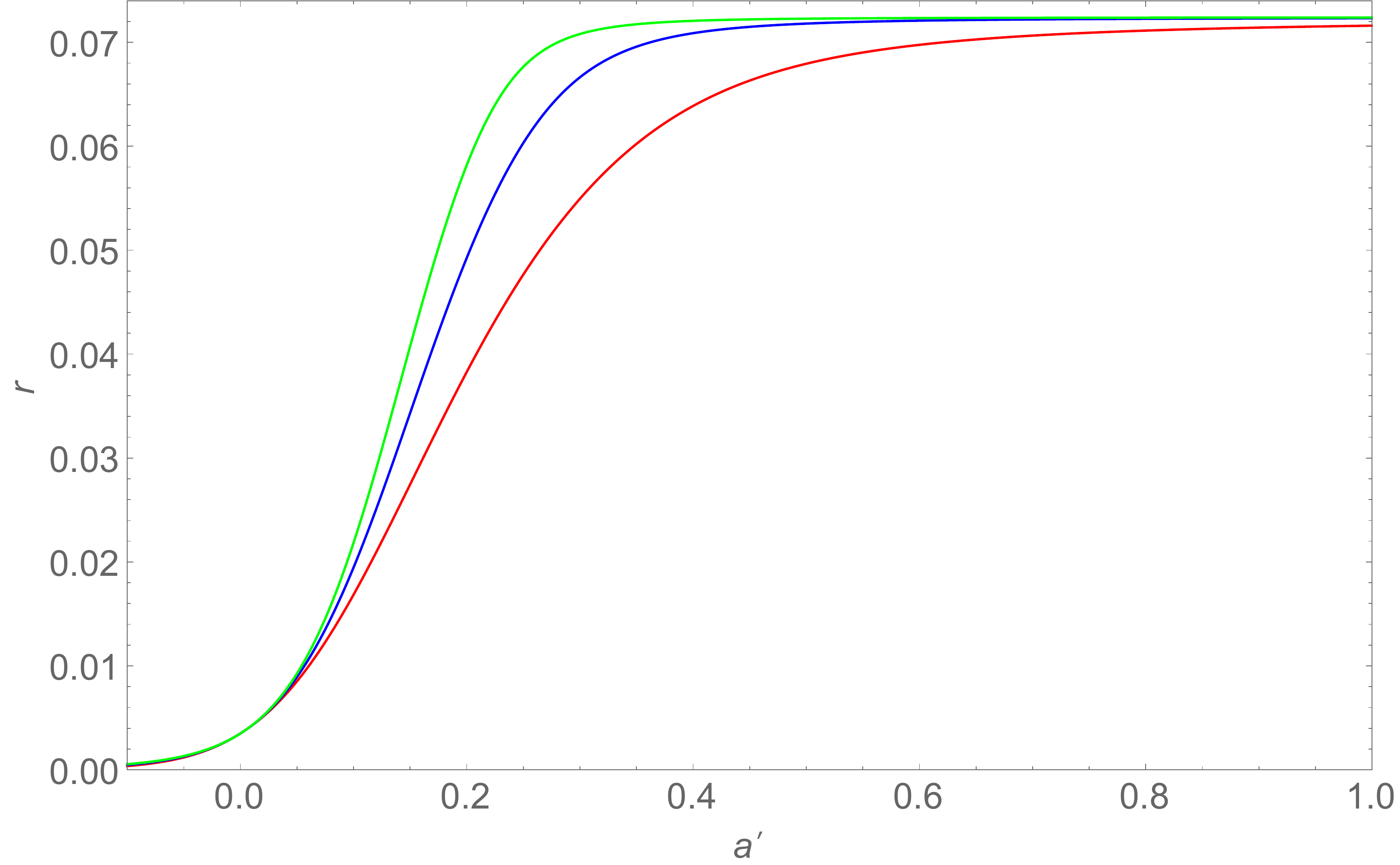}
\caption{$n_S$ vs $a^\prime$ (left) and $r$ vs $a^\prime$ (right) for $\xi = 100$ (red curve), $1000$ (blue curve), $5000$ (green curve). 
}
\label{nsandrversusalinha}
\end{figure}

The relation between the inflationary parameters and the radiative corrections can be better visualised in the Fig. \ref{nsandrversusalinha}, where we present the predictions of the model in the $a^\prime \times (n_S,\, r$) plane. In particular, $n_S$ achieve a maximum somewhere between $0.1 < a^\prime <0.2$, with a slightly different value for each curve. In the large correction regime (large $a^\prime$), all the three curves approach to a slightly close asymptotic value, around $n_S \simeq 0.973$. Meanwhile, the predicted tensor-to-scalar ratio $r$ runs from small values in the negative regions of $a^\prime$ to a asymptotic value $r \simeq 0.072$ at large radiative corrections. In both panels, the parameters seem to acquire a slight sensitivity to $\xi$ in the region $0.1 \lesssim a^\prime \lesssim 0.6$. For the region of the phase diagram that leads to $a^\prime < -0.07$ the predictions for $n_S$ run quickly to tiny values, raising a tension with Planck results. 


\begin{figure}[!h]
\centering
\includegraphics[scale=0.3]{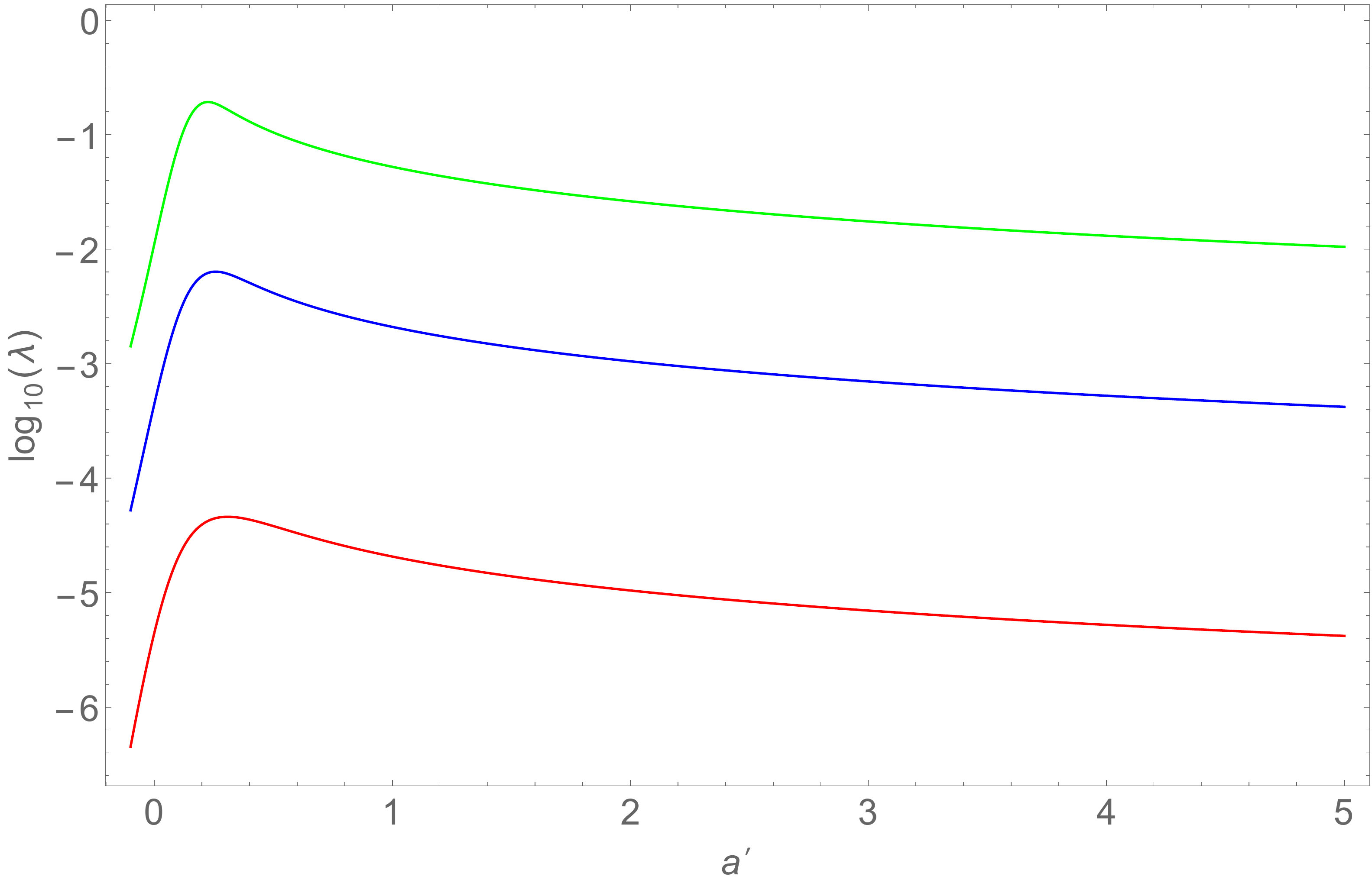}
\caption{$\log(\lambda)$ vs $a^\prime$ for $\xi = 100$, $1000$ and $5000$. Colors are labeled as in Fig. \ref{nsandrversusalinha}.
}
\label{Fig:al_lamb}
\end{figure}

One of the most revealing constraint of the model comes from the measured value of the amplitude of scalar perturbations, $A_S \simeq 2.1 \times 10^{-9}$ for the pivot choice $k_\ast=0.05$ Mpc$^{-1}$ \cite{Aghanim:2018eyx}. By inverting the expression resulting from \eqref{eq:PR} one can write the value of the quartic coupling $\lambda$ in terms of $\xi$ and $a^\prime$. In other words, the non-minimal coupling and the radiative corrections are degenerated in the value of the quartic Higgs coupling. In Fig. \ref{Fig:al_lamb} we present the graphic consequence of such a degeneracy. Although the form of the curves do not seem to significantly alter with variations in $\xi$, the magnitude of the amplitude $\lambda$ is highly dependent on the non-minimal coupling $\xi$. Furthermore, each curve exhibits a mild variation with $a^\prime$, presenting a maximum for $\lambda$ somewhere between $0.2 \lesssim a^\prime \lesssim 0.3$.

Fig. \ref{Fig:al_lamb} depicts the constraints on the potential amplitude resulting from the slow-roll analysis of inflation. At lower energy scales, however, the magnitude of $\lambda$ is fixed by the phenomenology of the Higgs boson. In the following section, we revisit the running equations for Higgs Inflation model in order to extrapolate the electroweak bounds on the Higgs quartic coupling to the inflationary energy regime. 





\section{Renormalization Group Equations}
\label{sec:RGE}
The observables of a renormalized field theory must not depend on the specific scale $\mu$ at which the Lagrangian parameters are defined. This gives rise to the running (flow) of the renormalized couplings between two distinct energy scales. Such relation is expressed mathematically by the chain rule \cite{Coleman:1973jx,Sher:1988mj},
\begin{equation}
    \left( \mu \frac{\partial}{\partial \mu} + \beta_i \frac{\partial}{\partial \lambda_i}- \gamma \frac{\partial}{\partial h}\right)V_{eff}=0,
\end{equation}
where
\begin{equation}
    \beta_i = \mu \frac{\partial \lambda_i}{\partial \mu}\,\,\,\, \text{and}\,\, \,\,\gamma = -\frac{\mu}{h}\frac{dh}{d\mu} \label{betas}
\end{equation}
are the $\beta$-functions for the SM couplings ($\lambda_i = {\lambda, y_t, g^\prime, g, g_S,...}$) and the anomalous dimension of the Higgs field, respectively. The set of equations in \eqref{betas} are often called renormalization group equations (RGE). They determine the flow of the Lagrangian parameters according to the symmetries and particle content of the model.


For the Higgs Inflation scenario, the non-minimal gravity sector manifests in the RGEs through the suppression of the off-shell Higgs propagator. There are two numerically equivalent ways to compute this suppression. The first one introduces a factor $s$ to each Higgs propagator\footnote{Only the radial mode of the Higgs doublet, $h$, is suppressed \cite{Bezrukov:2009db,DeSimone:2008ei,Lerner:2009xg}.}, with 
\begin{equation}
    s(h) =\frac{1+\xi h^2/M^2_P}{1+(1+6\xi)\xi h^2/M^2_P} \label{Suppress}
\end{equation}
following from the commutation relation for the Higgs field \cite{DeSimone:2008ei,Barvinsky:2009fy}. The second method imposes the abrupt suppression (freezing) of the radial mode of the Higgs doublet at scales of energy larger than $M_P/\xi$, with the purpose to attenuate its effective coupling to the other SM fields \cite{Bezrukov:2009db}. In this section, we follow closely the analysis performed in \cite{Allison:2013uaa}, where the two-loops $\beta$-functions for the non-minimal Standard Model were computed with the suppression factor $s$ (appendix \eqref{Sec:RGE}).



We are particularly interested in the evolution of the Higgs quartic coupling $\lambda$, the top quark Yukawa coupling $y_t$, the electroweak $g^\prime, g$ and strong $g_S$ gauge couplings, in order to compute the main contributions to $a^\prime$ at Planck scale. According to the standard procedure, we define the SM parameters at electroweak scale as contour conditions, in order to solve the set of coupled differential equations in \eqref{betas}. From the global fit of electroweak precision data \cite{Zyla:2020zbs}, we obtain the values for the $\overline{\text{MS}}$ gauge couplings $g$ and $g^\prime$,
\begin{eqnarray}
&& g(\mu = M_Z) = \sqrt{4\pi \alpha_{em}\sin^2{\theta_W}} = 0.651784, \\ 
&& g^\prime(\mu = M_Z) = \sqrt{4\pi \alpha_{em}\cos^2{\theta_W}} = 0.35744,
\end{eqnarray}
where $M_Z \simeq 91.19$ GeV is the $Z$ boson pole mass.

In what concerns $\lambda$, $y_t$ and $g_S$, higher order threshold corrections are supposed to contribute considerably at the weak scale, given the magnitude of QCD and top Yukawa interactions. {{In the following analysis we make use of the interpolating formul$\ae$ originally obtained in}} \cite{Buttazzo:2013uya},
\begin{eqnarray}
 \lambda (\mu=M_t) = 0.12604 + 0.00206 \left(\frac{M_H}{\text{GeV}}-125.15 \right) - 0.00004 \left(\frac{M_t}{\text{GeV}} - 173.34 \right), \\
 y_t (\mu=M_t) = 0.93690 + 0.00556\left(\frac{M_t}{\text{GeV}} - 173.34 \right) - 0.00042\frac{\alpha_S(M_Z)-0.1184}{0.0007}, \\
 g_S (\mu = M_t) = 1.1666 + 0.00314 \frac{\alpha_S(M_Z)-0.1184}{0.0007} - 0.00046\left(\frac{M_t}{\text{GeV}} - 173.34 \right),
\end{eqnarray}
where $\alpha_S(M_Z)$ is the $\overline{\text{MS}}$ strong coupling and $M_h,M_t$ are the pole masses of the Higgs boson and the top quark, respectively. The full two-loop threshold corrections were considered in order to obtain these expressions. Lastly, the value of the non-minimal coupling $\xi$ must be defined at some large energy scale. The analysis bellow is performed for $100 \leq \xi(M_P) \leq 5000$.

\begin{figure}[!h]
\centering
\includegraphics[scale=0.45]{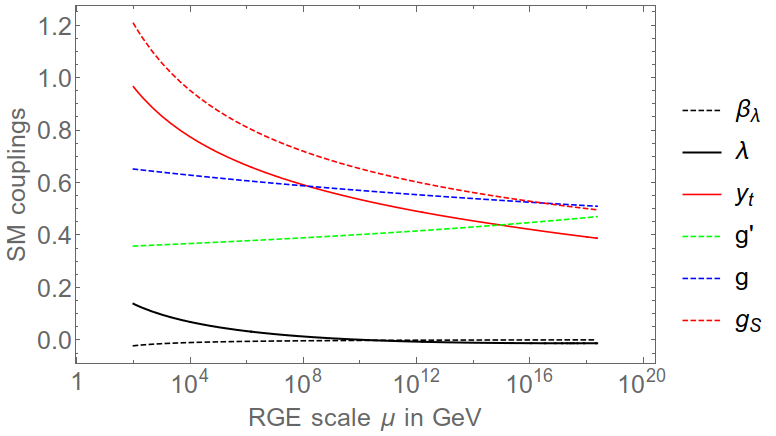}
\caption{Flow of the $\overline{\text{MS}}$ standard parameters $\lambda, y_t, g^\prime, g, g_S$ and $\beta_\lambda$ according to the renormalization scale $\mu$. In addition to the electroweak contour conditions discussed in the text, $\xi (M_P) = 1000$ was used in order to obtain the points.
}
\label{Fig:MS_MU}
\end{figure}

With the current global fit of the electroweak precision data and the two-loop $\beta$-functions (appendix \eqref{Sec:RGE}), one can solve the set of RGEs. Specifically, the central values of the observed strong gauge coupling $\alpha_S(M_Z) = 0.1179 \pm 0.0010$, Higgs pole mass $M_H = 125.10 \pm 0.14$ GeV and the top pole mass $172.76 \pm 0.30$ GeV \cite{Zyla:2020zbs} have been employed in order to obtain the flow of the SM parameters depicted in Fig. \ref{Fig:MS_MU}. As previously obtained in the literature \cite{Allison:2013uaa,Degrassi:2012ry}, the Higgs quartic coupling evolve to small negative values. In particular, the non-minimal coupling to gravity acts to suppress the Higgs interactions, lowering the instability scale of the Higgs potential to $\sim 10^{10}$ GeV. Also, note that the $\beta$-function for the Higgs quartic coupling remain small and roughly scale independent at high energies. Such behaviour substantiate the Coleman-Weinberg approximation used to describe the inflationary potential \eqref{veff}.

The results shown in Fig. \ref{Fig:MS_MU} only reinforce the well-known fact that the SM vacuum is metastable \cite{Espinosa:2007qp,Bezrukov:2012sa,Degrassi:2012ry,Buttazzo:2013uya,Bednyakov:2015sca}. In particular, the electroweak precision data favour an instability scale around $10^{10}-10^{11}$ GeV for the Higgs potential energy. Of course, this conjuncture compromises the viability of Higgs inflation. In the following analysis we choose to set all but the top quark pole mass to the central values of electroweak observations. More specifically, we use as input parameter $170 \leq M_t \leq 173$, in order to obtain the set of contour conditions for $\lambda, y_t$ and $g_S$. Such approach is justified given the inherent uncertainties in the definition of the Monte Carlo reconstruction of the top quark pole mass \cite{Butenschoen:2016lpz}. In addition, we set the non-minimal coupling $100 \leq \xi \leq 5000$, at energy scale $M_P$, following the previous slow-roll discussion.

\begin{figure}[!h]
\centering
\includegraphics[width=0.47\textwidth, height=0.2\textheight]{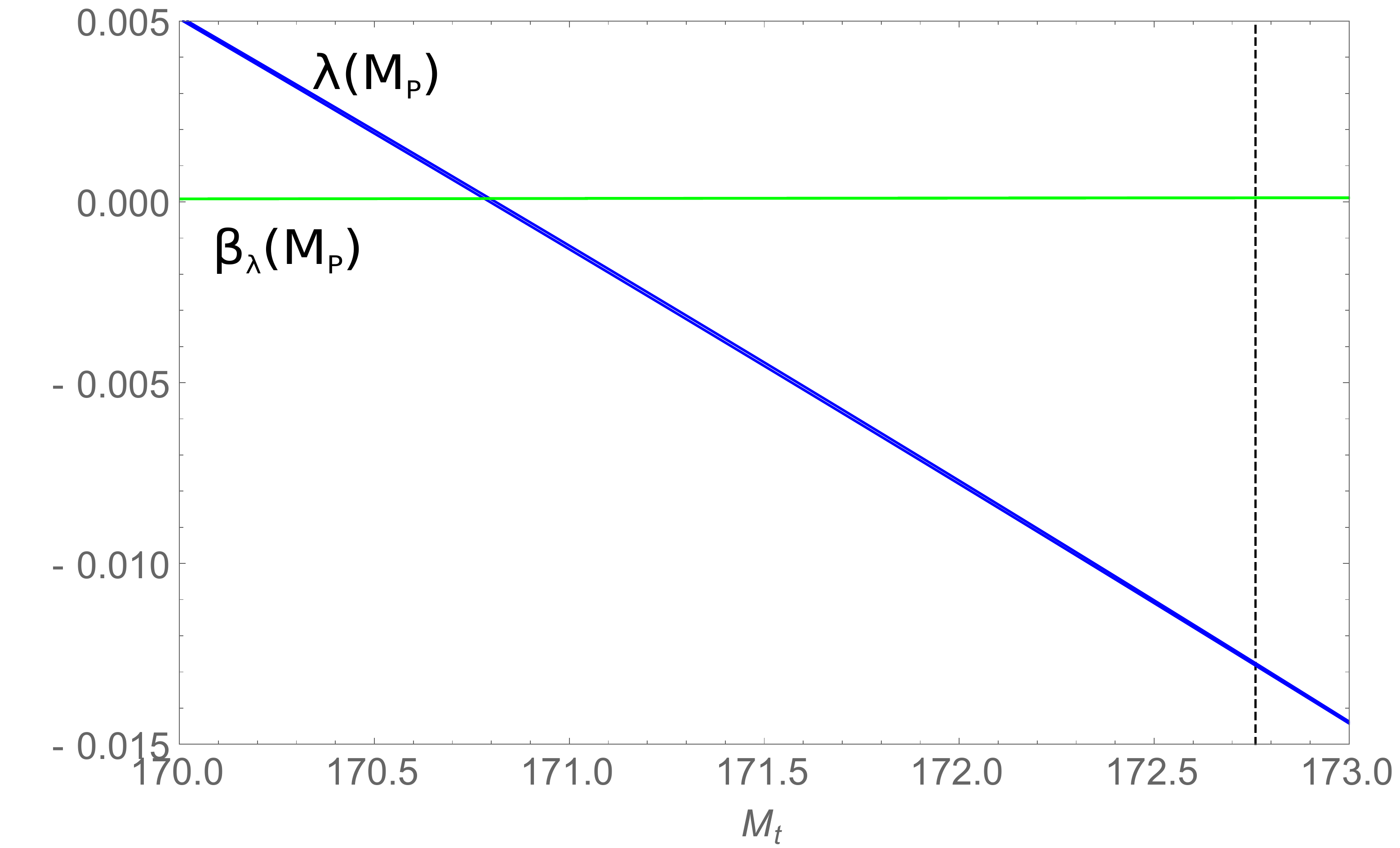}
\includegraphics[width=0.45\textwidth, height=0.195\textheight]{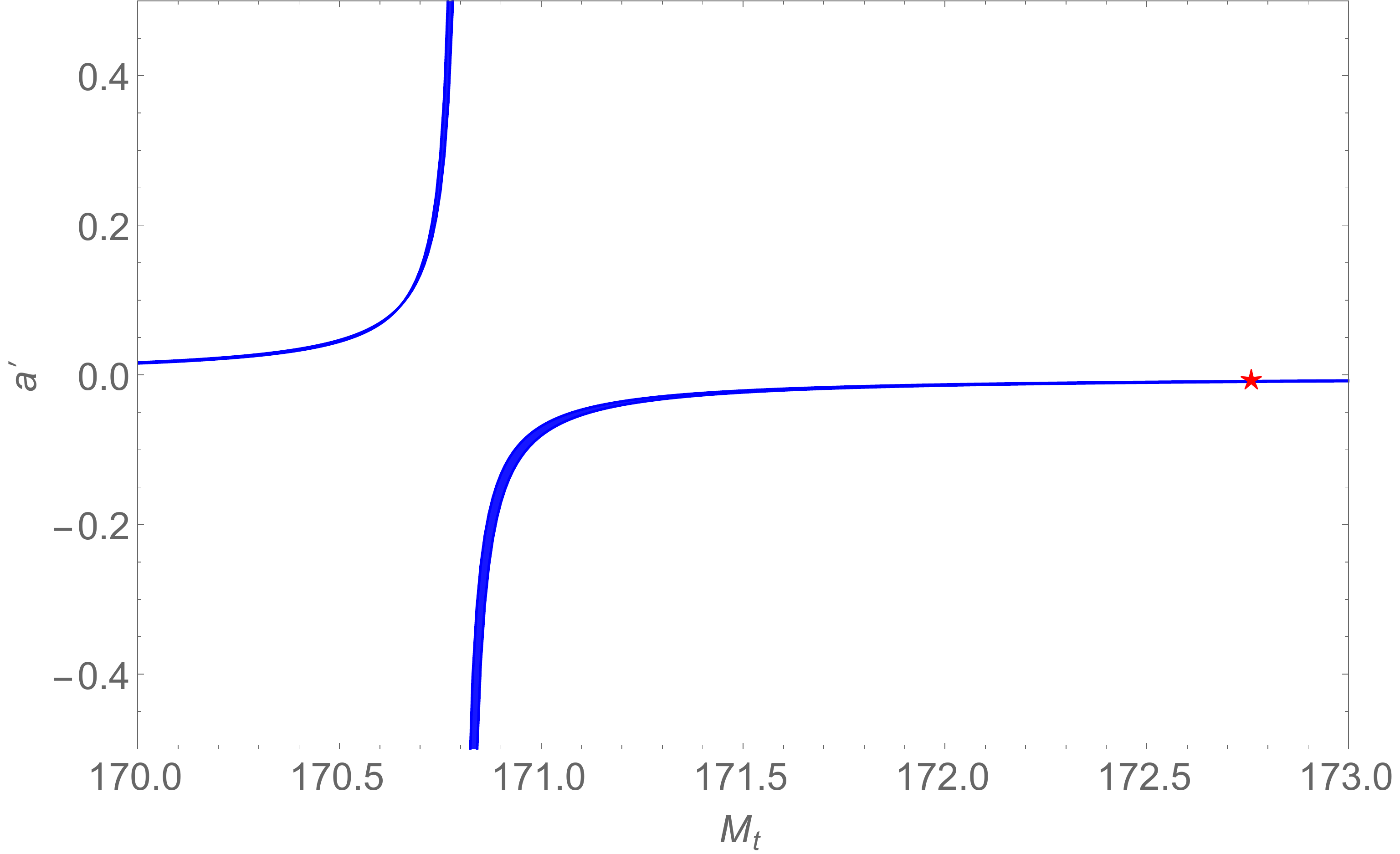}
\caption{$\lambda(M_P),\beta_\lambda(M_P)$ vs $M_t$ (left) and $a^\prime$ vs $M_t$ (right). The thickness in each curve reflects the variation in the non-minimal coupling $\xi$, defined at Planck scale, $100 \leq \xi(M_P) \leq 5000$. Also, the observed value of the top quark pole mass is indicated by the black dashed line (left) and the red star (right).
}
\label{Running_2}
\end{figure}

Fig. \ref{Running_2} presents the numerical dependence of the Higgs quartic coupling $\lambda$ and its one-loop $\beta$-function $\beta_\lambda$, computed at Planck scale, to the top quark pole mass $M_t$ (left panel). The numerical values for the corresponding $a^\prime$ parameter are also presented (right panel). For comparison purposes, the central value of the top quark pole mass $M_t = 172.76$ GeV \cite{Zyla:2020zbs} is marked by the black dashed line in the left panel, while the corresponding value of $a^\prime$ is indicated by the red star in the right panel. The thickness of each curve results from the variation of the non-minimal coupling constant, from $\xi = 100$ to $\xi =5000$, showing a mild dependence of the running equations to this parameter \cite{Steinwachs:2019hdr}. 
As expected, the magnitude of $\lambda(M_P)$ becomes greater and positive as $M_t$ moves away from its observed low-energy value. In particular, $\lambda (M_P) \simeq 0$ is obtained for $M_t$ somewhere close to $170.80$ GeV. The $\beta$-function for the Higgs quartic coupling remains small and positive for the entire range of $M_t$ considered. Consequently, the magnitude of $a^\prime$ is singular for $M_t \simeq 170.8$ GeV (according to its definition $a^\prime = \beta_\lambda/\lambda$).

On the other hand, the inflationary dynamics of the Higgs field can be used to constrain the parameter space of the SM in order to describe cosmological observations \cite{Aghanim:2018eyx}. In particular, the amplitude of scalar perturbations prescribe the relation between the Higgs quartic coupling and the radiative corrections to the Higgs potential, according to the discussion preceding Fig. \ref{Fig:al_lamb}. It is instructive to compare the constraints resulting from the low and high energy behaviour of the theory. 

\begin{figure}[!h]
\centering
\includegraphics[scale=0.28]{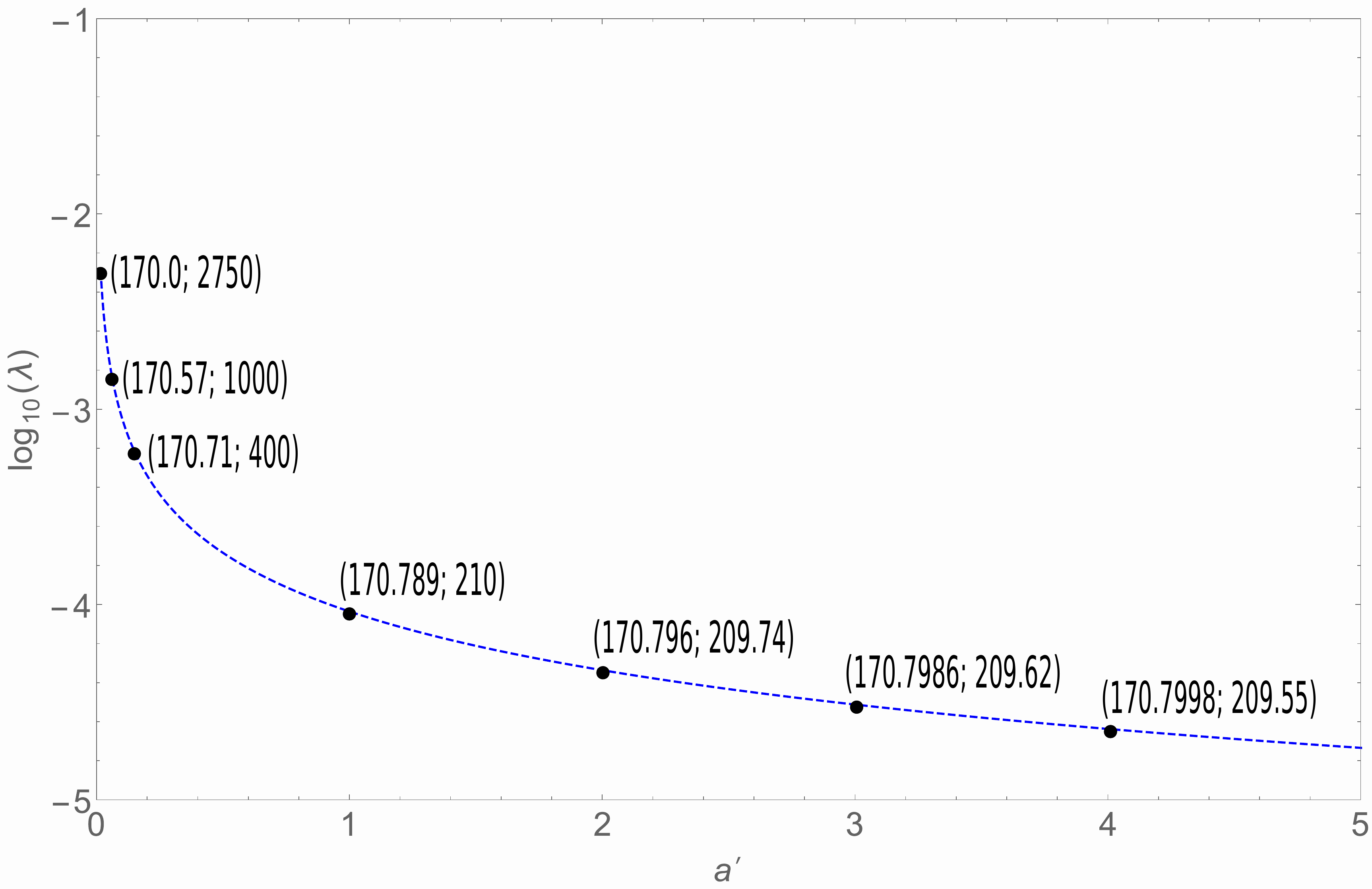}
\caption{$\log_{10}(\lambda)$ vs $a^\prime$. The blue dashed line is obtained for the solution of the RGEs with $M_t$ ranging from $170.00$ GeV (leftmost point) to $170.80$ (for the limit $a^\prime \rightarrow \infty$). Throughout the curve a series of benchmark points are highlighted, with the first number in parenthesis representing the value of the top quark pole mass $M_t$ employed to solve the threshold corrections in the electroweak scale parameters, while the second number represents the value of the high energy scale non-minimal coupling $\xi(M_P)$, required to recover the measured amplitude of scalar perturbations ($A_s = 2.101 \times 10^{-9}$ \cite{Aghanim:2018eyx}).
}
\label{fig:Running_3}
\end{figure}

In Fig. \ref{fig:Running_3} we present the results of such analysis in the $a^\prime \times \lambda$ plane. The dashed blue curve is obtained through the solution of the RGEs, where the top quark pole mass assumes the values between $170.00 \leq M_t \leq 170.80$ GeV. 
Some of the points are highlighted with black dots, with the input value for the low-energy quark mass displayed by the first number in parenthesis. Following the previous analysis, the Higgs quartic coupling goes asymptotically to zero as $M_t$ approaches $170.80$ GeV. While the running equations are nearly insensitive to variation in $\xi$, one can tune the non-minimal coupling to gravity in order to reproduce the observed value of the amplitude of scalar perturbations $A_s$. The resulting values for $\xi$ are displayed by the second number in parenthesis for each benchmark point in the figure. Following expression \eqref{eq:PR}, $\xi$ assumes increasingly smaller values as $\lambda$ becomes smaller. Negative values for the Higgs quartic coupling are not compatible with the observed value of $A_s$, enforcing the exclusion of the negative solutions to $\lambda(M_P)$.

Following Fig. \ref{fig:Running_3} one could be led to infer that increasingly lower values of $\xi$ could be imposed in order to fit the observed value of the amplitude of scalar perturbations for increasingly lower values of $\lambda$. However, for $\lambda \rightarrow 0$ the one-loop beta function for the quartic coupling reads $\beta_\lambda \sim 8.73 \times 10^{-5}$ (obtained for $M_t \simeq 170.80$ GeV). With these set of values, one has to assume $\xi \simeq 204.06$ in order to recover the observed value of $A_s$. This translates as a lower bound to the non-minimal coupling, \textit{i.e.}, $\xi \gtrsim 204.6$ for this approximation of the Higgs Inflation scenario.

The central discussion on the Higgs inflation scenario follows from the difficulty to reconciling inflationary dynamics with electroweak phenomenology.
As depicted in Fig. \ref{fig:Running_3}, the observed value of the amplitude of scalar perturbations diverge from the observed low-energy quark pole mass ($M_t = 172.76 \pm 0.30$ GeV \cite{Zyla:2020zbs}).
On the other hand, the purely slow-roll analysis of the empirical parameters of inflation ($A_s,\,\, n_S,\,\, r,...$) may lead to an inaccurate interpretation of the cosmological data, as pointed out in \cite{Mortonson:2010er} and demonstrated in \cite{Campista:2017ovq,Rodrigues:2020dod} for non-minimal models of inflation. In the next section, we intend to contribute to this discussion by performing a Monte Carlo Markov Chain analysis of the theory, confronting the model predictions with the most recent data.

\section{Analysis Method and Results}
\label{sec:Method&Results}
In this section we proceed with the discussion of the methodology used to compare our model with cosmological data. 
{We build our theory assuming a standard cosmological model with a modified primordial power spectrum, following our Sec. \ref{sec:Slow-Roll}. This means that we consider the usual cosmological parameters, namely, the physical baryon density, $\omega_b$, the physical cold dark matter density, $\omega_{cdm}$, the optical depth, $\tau$, and the angular
diameter distance at decoupling, $\theta$, while we do not consider the standard treatment of the primordial power spectrum, which assumes a power law parametrization with a primordial scalar amplitude, $A_s$, and a primordial spectral index, $n_s$. Instead of this assumption, we consider the primordial power spectrum given by the inflationary potential Eq.(\ref{eq:InfPot}), where we have two more parameters: the non-minimal coupling, $\xi$, and the radiative corrections at Planck scale, $a^\prime$.

We modify the numerical Code for Anisotropies in the Microwave Background ({\sc CAMB})~\cite{Lewis:1999bs} following the indications of~{\sc ModeCode}~\cite{Mortonson:2010er, Easther:2011yq}, which allows the primordial spectrum of an inflationary model to be analysed by giving a specific form of the potential V($\phi$) (for further details, see Ref.\cite{Rodrigues:2020dod}).
{\sc CAMB} is able to calculate the observational predictions of the model considered, thus allowing to compare the model with the data. It is implemented in the {\sc CosmoMC} packages ~\cite{Lewis:2002ah}, that is a Monte Carlo Markov chain (MCMC) code able to perform parameters estimation of the considered model, as well as statistical analysis.

Once we have implemented our model in the code, we can resume the previous discussion about the parameters $\lambda, \,\, a^\prime$ and $\xi$ to understand their effects on the primordial power spectrum, $P_R$, and select the parameters prior to use in our analysis (see details explained later). For the other cosmological parameters, broad and flat priors are used, as usual.
In the following, we will explore two numerical equivalent ways to probe the parameter space of the Higgs Inflation model, each of them allowing complementary interpretations for the cosmological data. 
We compare our modeling with a dataset that combines the CMB Planck (2018) likelihood \cite{Aghanim:2019ame}, using Plik temperature power spectrum, TT, and HFI polarization EE likelihood at $\ell \leq 29$; BICEP2 and Keck Array experiments B-mode polarization data \cite{Ade:2018gkx}; BAO measurements from 6dFGS ~\cite{Beutler:2011hx}, SDSS-MGS~\cite{Ross:2014qpa}, and BOSS DR12~\cite{Alam:2016hwk} surveys, and the Pantheon sample of Supernovae Type Ia \cite{Scolnic:2017caz}.
\subsection{Exploring the quark top pole mass, $M_t$}
\begin{table}[t!]
\centering
\begin{tabular}[t]{|c | c | c | c |}
$\xi$  & $M_t$ & $r_{02}$ & $H_0$ \\
\hline
$2000$  & $170.304 \pm 0.005$ & $0.00556 \pm 0.00003$ & $68.11 \pm 0.40$ \\
$1000$  & $170.575 \pm 0.002$ & $0.0103 \pm 0.0001$ & $68.72 \pm 0.39$  \\
$500$   & $170.683 \pm 0.001$ & $0.0225 \pm 0.0002$ & $69.01 \pm 0.41$ \\
$209.5$ & $170.772 \pm 0.004$ & $0.0665 \pm 0.0008$ & $68.11 \pm 0.40$
\end{tabular}
\caption{ }
\label{tab:Mt}
\end{table}
%
\begin{figure}[!t]
\centering
\includegraphics[scale=0.3]{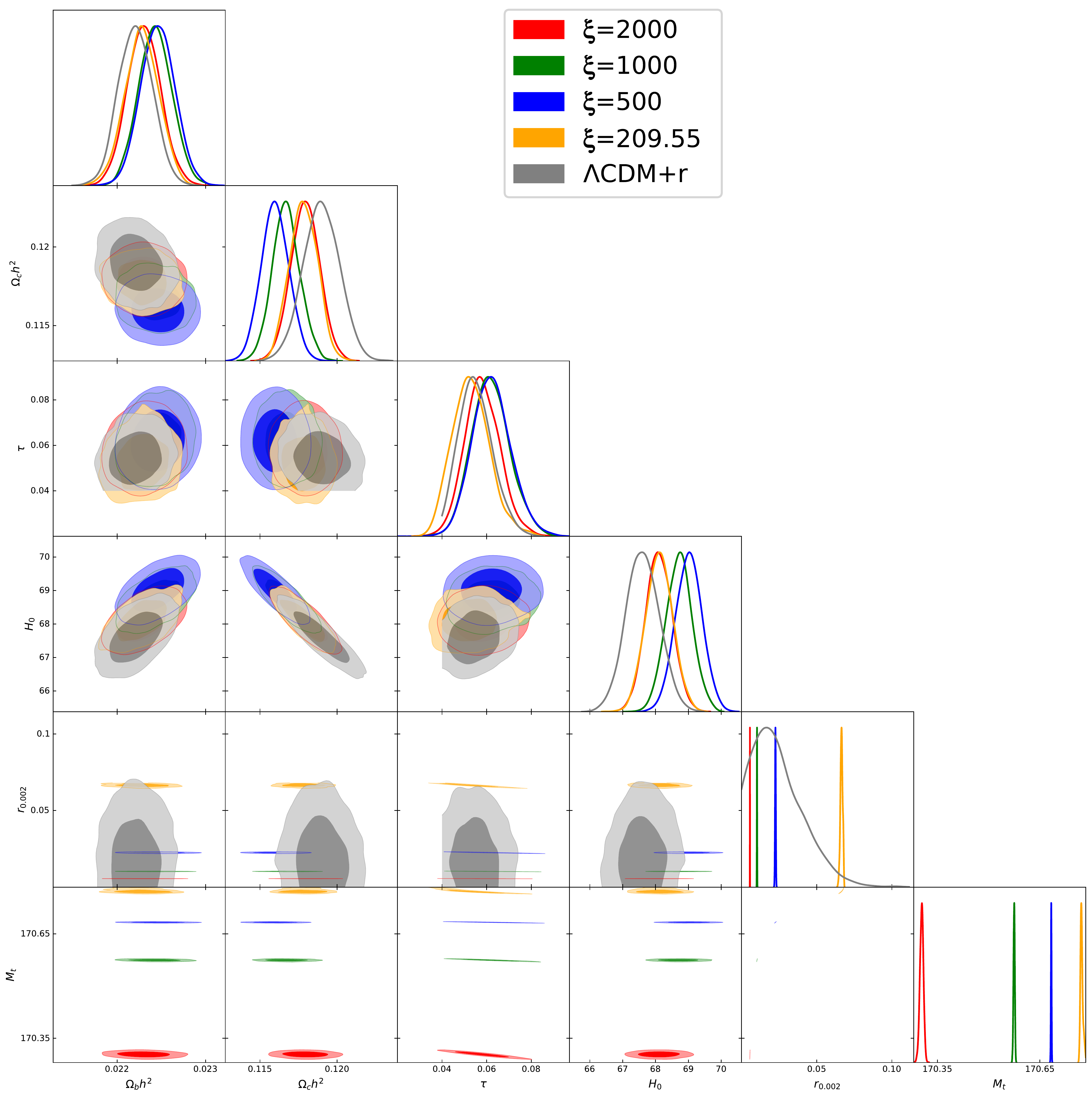}
\caption{2D C.L. and posterior distributions of SeeSaw and LCDM+r models}
\label{Fig:Analysis_Mt}
\end{figure}

{The first approach arises from the fact that the parameters composing the primordial potential Eq. \eqref{eq:InfPot} are related to the electroweak scale observables through the Renormalization Group Equations, described in appendix \eqref{Sec:RGE}. Following the discussion in Sec. \ref{sec:RGE} we set the prior to the quark top pole mass $170.00 \leq M_t \leq 170.79$ GeV \footnote{We have intentionally excluded the value $M_t \simeq 170.80$ in order to avoid the singularity $a^\prime \rightarrow \infty$.}. This in turn translates to the range $5.1 \times 10^{-3} \lesssim \lambda(M_P) \lesssim 9.0 \times 10^{-5}$ and $1.5 \times 10^{-2} \lesssim a^\prime(M_P) \lesssim 1.0$ for the Lagrangian parameters. Also, setting the value for the non-minimal coupling, we obtain the amplitude of scalar perturbations through expression Eq.\eqref{eq:PR}. In order to encompass the Planck's observed value for the primordial amplitude in our priors, we opt to select values of the non-minimal coupling parameter inside the interval allowed by the slow-roll analysis, $204.06 \lesssim \xi \lesssim 2750$ (see Fig. \eqref{fig:Running_3} and discussion). Therefore, we choose to perform the MCMC analysis for $\xi = 209.55,\,\, 500,\,\, 1000$ and $2000$. }

Our results are summarised in Fig.(\ref{Fig:Analysis_Mt}) and Tab. \ref{tab:Mt}, where we show the parameter constraint of our analysis using $\xi=2000$ (red lines), $\xi=1000$ (green lines), $\xi=500$ (blue lines) and $\xi=209.55$ (orange lines), compared to the $\Lambda$CDM+r model using the same dataset (grey line).
We note that for increasing coupling values we obtain lower $M_t$ values, going from $M_t= 170.772 \pm 0.004$ for $\xi=209.55$ to $M_t= 170.304 \pm 0.005$ for $\xi=2000$. This produces both a very precise expectation on the value of the tensor-to-scalar ratio, and a significant difference on the constraint of the expansion rate today, $H_0$. In particular, for decreasing coupling values the value of $H_0$ increases, which happens up to $\xi=500$. On the other hand, we note that the case $\xi=209.55$ shows constraints on  the cosmological parameters very close to the case with coupling $\xi=2000$. Considering a coupling of $\xi=500$, the discrepancy with the value estimated by the SH0ES Collaboration $H_0=74.03 \pm 1.42$ \cite{Riess:2019cxk} is lowered to $2.3 \sigma$, in comparison to the one obtained with the canonical $\Lambda$CDM parameterization ($4.4 \sigma$ \cite{Riess:2019cxk}). It is worth mentioning that the model investigated in this analysis uses five parameters, i.e, one less than the standard cosmological model, and that the parameter $M_t$ is closely related to the constraints on primordial amplitude. Such constraint is very accurate using CMB data, and for this reason we obtain a highly accurate constraint on $M_t$. Through expression Eq.\eqref{eq:PR} and the RGEs, one can translate the constraint on $M_t$ into $A_s$, obtaining a value close to the one pointed out by Planck collaboration, $A_s \simeq 2.1 \times 10^{-9}$, independently of the magnitude of $\xi$. This raises a second possibility to probe the parameter space of the Higgs Inflation model, as explained in the next approach.


\begin{figure}[!t]
\centering
\includegraphics[scale=0.5]{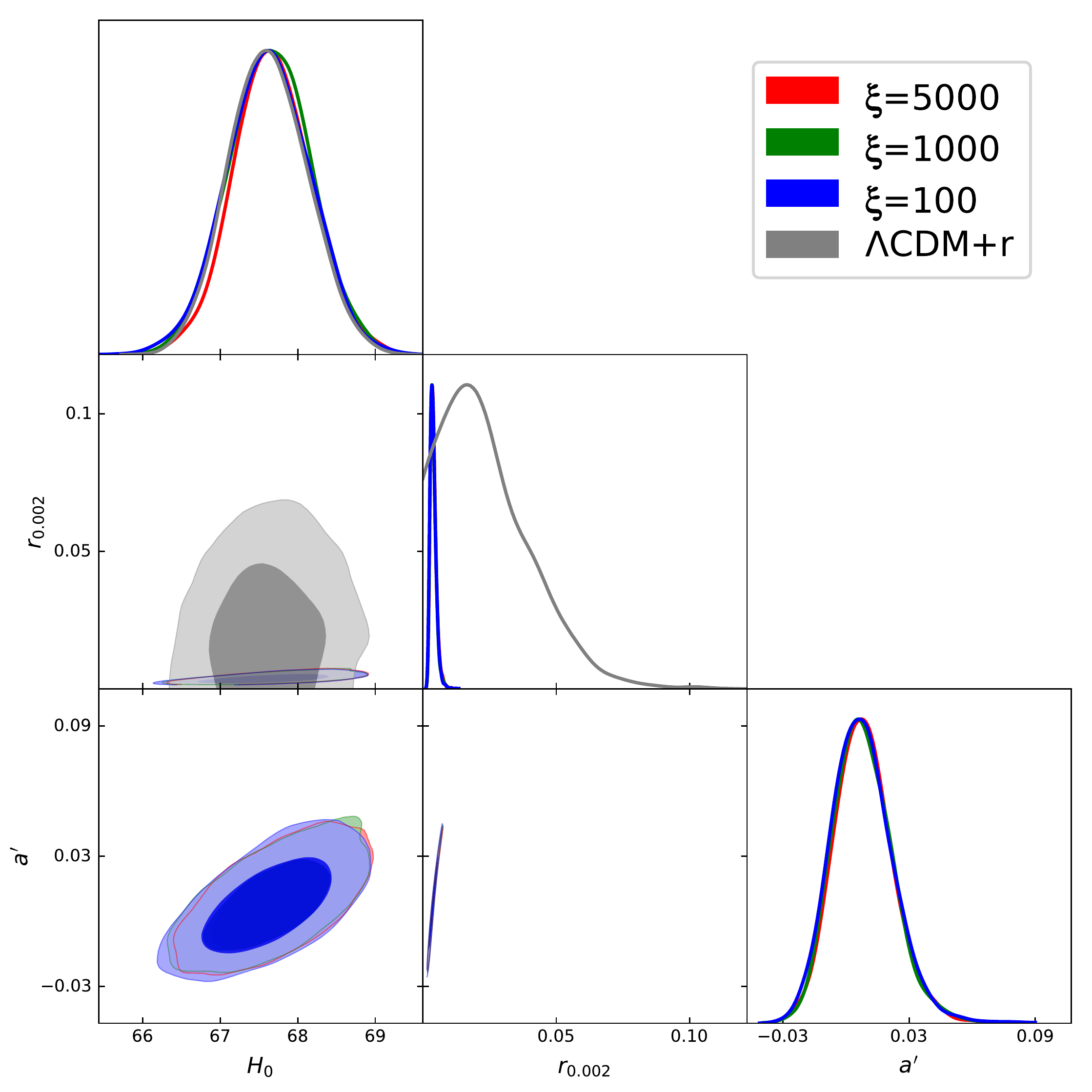}
\caption{2D C.L. and posterior distributions of SeeSaw and LCDM+r models}
\label{Fig:Analysis_a}
\end{figure}
\subsection{Exploring the radiative correction at Planck scale, $a^\prime$}

Following a procedure closely related to the one adopted in Ref. \cite{Rodrigues:2020dod} and the slow-roll discussion in Sec. \eqref{sec:Slow-Roll}, we fix the amplitude of scalar perturbations to the value compatible with the one inferred by Planck 2018 release, $A_s = 2.1 \times 10^{-9}$ \cite{Aghanim:2018eyx}. As result, we obtain the relation between $\lambda$, $\xi$ and $a^\prime$ depicted in Fig. \eqref{Fig:al_lamb}. Such relation reflects on the shape of the scalar potential and, consequently, on the prediction of the inflationary parameters (see Figs. \eqref{Fig:ns_r} and \eqref{nsandrversusalinha}). Similarly to the previous analysis, we can set the value for $\xi$ and employ the MCMC parameter estimation method to constrain $a^\prime$ and then link the constraint to $M_t$ by Renormalization Group methods\footnote{Remember that $a^\prime$ is closely related to the radiative corrections at Planck scale, $a^\prime=\beta_\lambda/\lambda$, which relates to the electroweak scale parameters through Renormalization Group equations.}. We will see in the following that this approach can lead to an unique constraint to $M_t$.

In this analyses, we choose a flat prior for $a^\prime$ in the range $[-0.05:4]$, looking at the Figs. (\ref{Fig:ns_r}-\ref{nsandrversusalinha}). We also fix $\xi$ to $100$, $1000$ and $5000$ in order to obtain a comprehensive understanding of this parameter. The results are summarized in Fig.(\ref{Fig:Analysis_a}), where we note that the constraints on the cosmological parameters overlap for each value explored for the non-minimal coupling. 
In particular, we obtain $a^\prime = 0.008  \pm 0.015$ at $68 \%$ C.L. for the data combination described above. In contrast to the previous analysis, where we derived the constraint on $M_t$ according with the value assumed for $\xi$, this approach results in a unique constraint on $a^\prime$, which can be translated to $M_t$ by means of Renormalization Group methods.

Following the discussion at Sec. \eqref{sec:RGE}, we set all but the top quark pole mass to the central values of the electroweak observations as contour conditions to the RGEs. After the numerical solution of the six coupled differential equations in appendix \eqref{Sec:RGE}, one can obtain the numerical value of $a^\prime$ substituting the solutions for $\lambda,\, y_t,\, g^\prime,\, g,\, g_S$ and $\xi$ into the definition of $a^\prime = \beta_\lambda/\lambda$.
Using this process with the upper one sigma value of $a^\prime$ constrained by our previous analysis, we find the upper limit to the top quark pole mass, $M_t \leq 170.222$ GeV. 
In other words, it would be required for the top quark pole mass to assume a value $\approx 8\sigma$ away from the one observed at low-energy experiments, $M_t = 172.76 \pm 0.30$ GeV \cite{Zyla:2020zbs}, in order the reconcile the low-energy phenomenology and inflationary dynamics of the Higgs field.

\section{Final Remarks}
\label{sec:Conclusions}
In this work we investigated the observational viability of a inflationary scenario driven by the 
Higgs boson, as proposed by Bezrukov and Shaposhnikov \cite{Bezrukov:2007ep}. We considered the Coleman-Weinberg approximation to the effective Higgs potential in order to obtain the predictions of the primordial power spectrum, and constrained the Lagrangian parameters at the inflationary scale.  Finally, through the solution of the two-loop Renormalization Group equation (RGE) for the Standard Model couplings, we compared our results with the electroweak scale observations \cite{Zyla:2020zbs}.

In order to explore the parameter space of the Higgs Inflationary model we made use of two numerical equivalent approaches, obtaining complementary interpretation of the data. First we solved the RGEs, with the electroweak scale value to the Standard Model observables as contour conditions, in order to obtain the values of the Lagrangian parameters at the inflationary scale, namely $\lambda$ and $a^\prime$. Fixing the value of the non-minimal coupling to gravity $\xi$, we then estimated the best-fit values to the cosmological parameters, as well as the quark top pole mass, through the MCMC analysis described above. In particular, $M_t$ is closely related to the amplitude of primordial perturbations, allowing a highly accurate constraint on this parameter. We obtained values for $M_t$ ranging from $170.772 \pm 0.004$ for $\xi=209.55$ to $170.304 \pm 0.005$ for $\xi=2000$. In what concerns the Hubble constant, the MCMC analysis revealed an apparent non-linear dependence between $H_0$ and $\xi$. For $\xi=500$ we obtained $H_0 = 69.01 \pm 0.41$ km/s/Mpc, lowering the tension with the value estimated by the SH0ES Collaboration $H_0=74.03 \pm 1.42$ to $2.3 \sigma$ \cite{Riess:2019cxk}.

On the second approach, we fixed the amplitude of scalar perturbations to the value indicated by Planck measurements, $A_s = 2.101 \times 10^{-9}$ \cite{Aghanim:2018eyx}. As a consequence, we obtained a relation between $\lambda$, $\xi$ and $a^\prime$ at high energy scales. This allowed us to investigate the impact of the radiative corrections to the Higgs potential $a^\prime$ in the primordial power spectrum. For a wide range of the non-minimal coupling to gravity $\xi$, the MCMC analysis resulted in the same amount for the radiative corrections, $a^\prime = 0.008 \pm 0.015$ at $68\%$ (C.L.). One particularly useful aspect of this approach is that it leads to a unique limit to the quark top pole mass. In order to obtain $a^\prime(M_P)$ in the range suggested by the MCMC analysis one must assume $M_t \leq 170.222$ GeV in the solutions of the RGEs. This is a remarkable result since the value for the top quark mass is measured to be $M_t = 172.76 \pm 0.30$ GeV by the electroweak energy scale experiments \cite{Zyla:2020zbs}, resulting in a discrepancy of approximately $8\sigma$ between the observed low-energy value and the amount inferred in our MCMC analysis.

Our results are in agreement with previous works \cite{Bezrukov:2012sa,Allison:2013uaa,Masina:2018ejw}, and allows us to conclude that the simplest realisation of the Higgs Inflation model \cite{Bezrukov:2007ep} faces strong difficulties in reconciling cosmological data \cite{Aghanim:2019ame,Ade:2018gkx,Beutler:2011hx,Ross:2014qpa,Alam:2016hwk,Scolnic:2017caz} to the observations at electroweak scale \cite{Zyla:2020zbs}. On the other side, small deviations on the scenario proposed in \cite{Bezrukov:2007ep} could present viable candidates to explain the early universe dynamics and longstanding problems in the fundamental particle physics \cite{Lerner:2009xg,Okada:2011en,Ballesteros:2016euj,Ferreira:2017ynu,Rodrigues:2018jpv,Kawai:2020fjt}. Some of these analyses are currently in progress and will be reported in a forthcoming communication. Finally, given the sensitivity with which $M_t$ determines the expected value of $r$, we expect that future experiments on the polarised CMB signal, such as COrE \cite{DiValentino:2016foa}, Simons Observatory \cite{Abitbol:2020fvn,Salatino:2018voz} and CMB-S4 \cite{Abazajian:2019tiv, Abazajian:2020dmr}, to be of great relevance to this type of analysis.

\section*{Acknowledgments}

JGR acknowledges Conselho Nacional de Desenvolvimento Cient\'{\i}fico e Tecnol\'ogico (CNPq) for the financial support (grant no.~150343/2020-5). MB thanks support of the Istituto Nazionale di Fisica Nucleare (INFN), sezione di Napoli, iniziative specifiche QGSKY. JSA acknowledges support from CNPq (grants no.~310790/2014-0 and 400471/2014-0) and FAPERJ (grant no.~204282). The authors thank the use of CAMB and CosmoMC codes. We also acknowledge the use of the Observat\'orio Nacional for providing the computational facilities to perform our analysis.

\appendix

\section{Renormalization Group Equations for Higgs Inflation} \label{Sec:RGE}

In this appendix we list the RGE employed in our analysis, first derived in \cite{Allison:2013uaa}. In particular, the flow for the couplings $\lambda,\, y_t,\, g^\prime,\, g,\, g_S$ and $\xi$ were computed in the $\overline{\text{MS}}$ scheme at two-loop level. For each coupling we write $dx/dt=\beta_x$, where $t=\ln{(\mu/\mu_0)}$. The anomalous dimension $\gamma$, employed in the field reescaling \eqref{eq:FReescaling}, is also given. Also, note the presence of the suppression factor $s=s(h)$, as discussed in Sec. \ref{sec:RGE}, for each off-shell Higgs propagator.

The two-loop RGE for the Higgs quartic coupling reads as follow,
\begin{eqnarray}
\beta_\lambda &=& \frac{1}{(4\pi)^2}\left[ (18s^2+6)\lambda^2 - 6y^4_t +\frac{3}{8}(2g^4+(g^2+g^{\prime 2})^2)+(-9g^2 - 3g^{\prime 2} +12y^2_t) \lambda \right) \nonumber \\
&+& \frac{1}{(4\pi)^4}\left( \frac{1}{48}\left( (3s + 912)g^6 - (290-s)g^4g^{\prime 2} - (560 - s)g^2g^{\prime 4} - (380 - s)g^{\prime 6} \right) \right. \nonumber \\
&+& (38 - 8s)y^6_t - y^4_t\left(\frac{8}{3}g^{\prime 2} + 32g^2_S + (12 - 117s + 108s^2)\lambda\right) \nonumber \\ 
&+& \lambda \left(-\frac{1}{8}(181 + 54s - 162s^2)g^4 + \frac{1}{4}(3-18s+54s^2)g^2g^{\prime 2} + \frac{1}{24}(90+377s + 162s^2)g^{\prime 4} \right. \nonumber \\
&+& \left. (27 + 54s + 27s^2)g^2 \lambda + (9 + 18s + 9s^2)g^{\prime 2}\lambda - (48 + 288s - 324s^2 + 624s^3 - 324s^4)\lambda^2 \right) \nonumber \\
&+& \left. y^2_t \left(-\frac{9}{4}g^4 + \frac{21}{2}g^2g^{\prime 2} - \frac{19}{4}g^{\prime 4} + \lambda (\frac{45}{2}g^2 + \frac{85}{6}g^{\prime 2} + 80 g^2_S - (36 + 108s^2) \lambda) \right) \right].
\end{eqnarray}
For the top Yukawa coupling we have,
\begin{eqnarray}
\beta_{y_t} &=& \frac{y_t}{(4\pi)^2}\left[  \left( \frac{23}{6} + \frac{2s}{3} \right)y^2_t - 8g^2_S - \frac{17}{12}g^{\prime 2} - \frac{9}{4} g^2 \right] \nonumber \\
&+& \frac{y_t}{(4\pi)^4}\left[ -\frac{23}{4}g^4 - \frac{3}{4}g^2 g^{\prime 2} + \frac{1187}{216}g^{\prime 4} + 9g^2 g^2_S + \frac{19}{9} g^{\prime 2}g^2_S - 108g^4_S \right. \nonumber \\
&+& \left. \left( \frac{225}{16}g^2 + \frac{131}{16} g^{\prime 2} + 36g^2_S \right)s y^2_t + 6\left( -2s^2y^4_t - 2s^3y^2_t \lambda + s^2 \lambda^2 \right) \right].
\end{eqnarray}
Also, the running for the gauge coupling assume the form,
\begin{eqnarray}
\beta_{g^\prime} &=& \frac{g^{\prime 3}}{(4\pi)^2} \left[ \frac{81 + s}{12}   \right] + \frac{g^{\prime 3}}{(4\pi)^4} \left[ \frac{199}{18}g^{\prime 2} + \frac{9}{2} g^2 + \frac{44}{3}g^2_S - \frac{17}{6}s y^2_t \right], \nonumber \\
\beta_{g} &=& \frac{g^{3}}{(4\pi)^2} \left[ -\frac{39 - s}{12}   \right] + \frac{g^{3}}{(4\pi)^4} \left[ \frac{3}{2}g^{\prime 2} + \frac{35}{6} g^2 + 12g^2_S - \frac{3}{2}s y^2_t \right], \nonumber \\
\beta_{g_S} &=& \frac{g^{3}_S}{(4\pi)^2} \left[ -7   \right] + \frac{g^{3}_S}{(4\pi)^4} \left[ \frac{11}{6}g^{\prime 2} + \frac{9}{2} g^2 - 26g^2_S - 2s y^2_t \right].
\end{eqnarray}
For the non-minimal coupling $\xi$,
\begin{eqnarray}
\beta_{\xi} &=& \frac{1}{(4\pi)^2}\left( \xi + \frac{1}{6} \right) \left[ -\frac{3}{2}g^{\prime 2} - \frac{9}{2}g^{2} + 6y^2_t + (6+6s)\lambda  \right] \nonumber \\
&+& \frac{1}{(4\pi)^4} \left( \xi + \frac{1}{6} \right) \left[ \left(-\frac{199}{16} + \frac{27}{8}s \right)g^4 + \left(-\frac{3}{8} + \frac{9}{4}s \right)g^2g^{\prime 2} + \left(\frac{3}{2} + \frac{485}{48}s \right)g^{\prime 4} \right. \nonumber \\
&+& \left( \frac{85}{12}g^{\prime 2} + \frac{45}{4}g^2 + 40 g^2_S \right)y^2_t + \left( 18 - \frac{63}{2}s \right)y^4_t + \left( 36g^2 + 12g^{\prime 2} - 36y^2_t \right)(1+s)\lambda \nonumber \\
&+&\left. \left( -108 + 126s -144s^2 + 66s^3 \right)\lambda^2 \right].
\end{eqnarray}
Finally, the Higgs anomalous dimension is given by,
\begin{eqnarray}
\gamma &=& -\frac{1}{(4\pi)^2}\left[ \frac{9}{4}g^2 + \frac{3}{4}g^{\prime 2} - 3y^2_t \right] \nonumber \\
&-& \frac{1}{(4\pi)^4} \left[ \frac{271}{32}g^4 - \frac{9}{16}g^2g^{\prime 2} - \frac{431}{96}sg^{\prime 4} - \frac{5}{2}\left( \frac{9}{4}g^2 + \frac{17}{12}g^{\prime 2} + 8g^2_S \right)y^2_t + \frac{27}{4}sy^4_t - 6s^3\lambda^2 \right].\nonumber \\
\end{eqnarray}

\bibliographystyle{JHEP}
\bibliography{bibliografia}

\end{document}